\documentclass{article}
\usepackage[utf8]{inputenc}
\usepackage{graphicx}
\title{Trust levels in social networks}
\author{Santanu Acharjee$^{1,\dagger}$, Akhil Thomas Panicker$^{2}$\\$^{1}$Department of Mathematics \\ Gauhati University, Assam, India\\
$^{2}$Department of Physics\\ Cochin University of
Science and Technology, Kerala, India\\
e-mails: $^{1}$scharjee326@gmail.com, $^{2}$atpanicker95@gmail.com\\
$^\dagger$Corresponding author: Santanu Acharjee}

\date{}

\begin{document}
\maketitle
{\bf Abstract:}
Dunbar's number is the cognitive limit of an
individual to maintain stable relationship with others in his network. It is based on the size of the neuro-cortex of human brain.  This number is found to be 150 and it is one of the main factors to maintain stabilities in  several types of  social networks. Recently,  Acharjee et al. [S. Acharjee, B. Bora, R.I.M. Dunbar. 2020. On M-Polynomials of Dunbar Graphs in Social Networks, \textit{Symmetry, 12(6)}, { 932}] proved that the small-world effects are only possible if everyone uses the full range of their network when selecting steps in the small-world chain. Trust is one of the major issues for one while selecting members form his social network. Moreover, literature of social networks, complexity, evolutionary biology, evolutionary psychology, etc. have proven that trust is one of  major factors for the evolution of one's social network with time. Most of the  literature on trust, Dunbar's number and social networks suggest that trust and Dunbar's number are interconnected in case of one's stable social network and trust needs time to be built after several social interactions, intimacy, etc. But, no literature has focused on the nature of one's trust levels on his acquaintances under Dunbar's number once  his network size starts to increase.  Do trust levels remain same for the individuals from one's perspective in his social network when the network size increases? It is one of the open questions in the  research area of complexity and trust based social networks  with Dunbar's number. In this paper, we find answer of  this important question regarding trust and Dunbar's number for one's social network by simulations. We also establish relation between  the power-law exponents $\alpha$ and trust cutoffs. Moreover, we also find that trust level never helps to diffuse information quickly or vice-versa to reach Dunbar’s number 150 along with hierarchy layers of 5, 15 and 50 individuals in networks of different sizes. At the end, we provide a conjecture about deception in social networks.\\

{\bf 2020 AMS Classifications:} 05C82; 54H30; 91D30; 94C15; 62P25; 05C82; 68Q06.\\  


\section{Introduction}

``Social networks"- it is an area of interdisciplinary research which was developed from the ideas of graph theory \cite {62} in last century. Indirectly to social networks,  Erd\H{o}s and R\H{e}nyi \cite{65} laid theoretical foundation of random graphs in 1960 and since then, it was believed that the social networks of individuals  were nothing but random graphs \cite{50}. After three decades, Barab\'{a}si and Albert \cite{50} disproved the existence of random graphs in several online networks, along with biological networks, etc. After Barab\'{a}si and Albert \cite{50}, social networks has grown to an area of indisciplinary area. It's important to note that pioneering mathematician Frank Harary realized importance of social networks several decades ago and thus made several contributions along with his collaborators in \cite{63,64} and many others. On the other hand, social networks is one of the interesting area for topologists. It is easy to find connection between topology and social networks. One may refer to  recent  research works \cite{52,53,54,55,56,57,58,59} for topological aspects of social networks. For connections of social networks with computer science, \cite{66,67} can be  refereed.\\

Trust helps to grow human social relationships [1]. It binds people by  benevolent acts and reciprocal interactions, and thus it becomes the basis of friendship [1,2]. According to Mitchell [3], trust is governed by social exchanges viz. social presence, verbal and nonverbal communication.  In these types of social exchanges, honesty or deception of the counter-party is assessed. A psychological definition of trust can be found in Rotter [4]. Rotter [4] defined trust as ``{\it an expectancy held by an individual or a group that the word, promise, verbal or written statement of another individual or group can be relied on}". Similarly psychological definition of deception is stated as ``{\it A successful or unsuccessful deliberate attempt, without forewarning, to create in another a belief that the communicator considers to be untrue in order to increase the communicator's payoff at the expanse of the other side}" [5].\\

Several theories on human friendship have suggested importance of reciprocity and exchange of benevolent acts in case of building social relationships [1,6,7]. Trust is one of the implicit factors for social relationships [1,6,7]. Experimental studies of Hays [6] and Oswald et al.[7]  showed the advantages of investing resources
on  fewer but more intimate social relationships. Moreover,
Gronovetter [8] showed that weak ties often helps one to get more information and resources. According to Baumeister and Leary [9], mutual trust starts to  grow between individuals, which leads to reliance on each other for emotional support, companionship, help, etc. when collaborations start to grow between groups. Trust
evolves to enhance collaboration on assessing quality of
trustworthiness or deception [10]. Hays [11] observed that close friendships have more weekly interactions, across a wider range of days, locations and times than that of casual friendships. It was also found that close friends offered more informal and moral supports than casual friends. Oswald et al. [7] classified different dimensions for trust-oriented behaviours in friendships viz. positivity, openness, interaction and supportiveness.
According to Sutcliffe et al. [12], best-friendships were found to be self-maintained.  With the help of computer simulations, Roberts and Renwick [13] showed that  reputations of individual on histories of his  collaboration help to build his social relationships.  Nowak and  Sigmund [14] established that cooperative strategies spread in populations if histories of
interactions are assessable in case of computer simulated repeated prisoner's dilemma. In case of sociocognitive models of trust, several factors influence trusting relationships [15,16]. Falcone and Castelfranchi [15,16]  identified degree of delegation between two parties, risks, the motivations and the goals shared by parties for social relationships, reputations as the factors to influence trusting relationships.  One may go through Castelfranchi and Falcone [17] more elaboration on social trust.
Moreover, it is found that indirect reciprocity helps to
form groups  and encourages social relationships [18].\\

There are several researches have done on trust from the
computational aspects. Sutcliffe and Wang [19] described the process of development of social relationships by a computational model. A several computational models have been established with trust.  Sutcliffe et al. [2] presented social trust model to investigate the relationship of social networks and social relationships in real and in social media. They found the existence of a few strong relationships in multilevel social structures. Their results demonstrated the existence of  more medium ties and large number of weak ties if one rewarded well-being and alliances with social interactions of high level.
Interested reader may refer [2] for extensive literature on trust and related computational models. For recent research in trust in case of Facebook and Twitter, one may refer to Sabatini and Sarracino \cite{68} and many others.\\

Both trust and deception are topics of equal interest to social scientists, economists and  network scientists. Gneezy [20] discussed the role of consequences though deception in economics. According to Gneezy [20], most of the economic interactions include deceptions.  Extreme condition of economic theory assumes each agent as ``homo  economicus" during economic interactions. It
means that each agent acts selfishly during economic interactions and does not bother about the well being of others. Thus, deception is one of the major properties for selfish acts of an economic agent. For eg., Akerlof [21] considered that that car sellers would always lie for their benefits in case  of asymmetric information. A similar argument in mechanism design  [22] says that people tell trust if there are scopes for  material  outcomes  as incentives. From the perspective of utilitarian
 philosophy, Bentham [23]  argued that an individual must weigh  benefits against harm, happiness against unhappiness, etc. before  telling a lie. But, a  contradictory view to utilitarian philosophy came from Bok  [24]. According to Bok [24],  there is no difference between a  truthful statement and a lie if both achieve same monetary  payoffs. Thus from [23] and [24], can we claim that we must be  cautious to trust individuals of our social networks?\\

There are four types of lying [20, 25]. They are (i) pro-social (ii) self-enhancement (iii) selfish (iv) antisocial. Gneezy [20] suggested that people consider gain at the time of telling lie. Moreover, people also care about loses of others along with own gains. Serota et al. [26] conducted an experiment with 1000 Americans and it was found that average  1.65 lies were told by each American per day. Similarly, other experiments [27-30] proved
that lies were told between frequency 0.6-2.0 per day, where face-to-face intersections  had less chance of deception in comparison to telephonic conversation.  Gino and Pierce [31] conducted experiments and found that people opted for dishonest behavior to relieve emotional distress which caused from wealth-based inequity. Moreover, people increased hurting behavior and reduced helping behavior in case of their own experiences to negative inequity or negative emotions and they also increased dishonest helping behavior when they experienced positive inequity or positive emotions [31]. Arnaboldi et al. [32] modelled the
trusted information diffusion pattern  in online social networks with an assumption that two nodes had trusted tie with reciprocal behavior. They [32] also showed that up to only $3\%$ of ties were used in case of information diffusion if there were strong ties in online social networks. They [32] also found that in case trustworthy paths for communications in online social networks,
two individuals were more than twice as far  away from each other.  Wang et al. [33] experimentally proved that badness
influences psychological reactions more than goodness, but
goodness influences more in behavior badness. Similar results were predicted by Panasiti et al. [34]. In case of deception, one's own reputation encourages other individuals of his social networks to be honest [34]. Moreover, they found that unfavorable situations
motivated  one to deceive in stronger way. Gino et al. [35]
conducted three experiments and concluded that individuals cheated more when their lies benefitted others and when the number of beneficiaries from their lies increased. There are several studies can be found on trust and deception. Trust and deception have very strong connection with individual's social networks in terms of Dunbar's number.\\

Social networks has mainly two approaches to be studied. One is ``six degrees of separation" or ``small world phenomenon [36,37] and the other one is Dunbar's number [38]. Six degrees of separation focuses on the maximum number of intermediate acquaintance links are needed to connect any two persons. On the other hand, Dunbar's number [38] is based on human cognition limit to maintain stable relationships with individuals in one's social network. Dunbar's number is one of the fundamental factors of
ego-centric social networks. Based on the correlative studies of the size of social networks and  neuro-cortex sizes of brains of primates and humans, Dunbar [38] found that an individual can maintain stable relationships with only 150 individuals of his social networks. This was previously found  for offline social networks,  but with  the evolution of internet, the existence of same result was confirmed in case of online social networks [39,
40]. Dunbar's hierarchy layers consist of alters with scaling ratio of 3 [41, 42].  These hierarchy layers consist of  a series of concentric circles of acquaintanceship [40, 43]. An individual is placed at the centre of these hierarchy layers and  remaining
concentric circles consist of  5, 15, 50, 150, 500, 1500, 5000 individuals cumulatively [40, 43]. The hierarchy layers consist of 5 and 15 individuals are called support cliques and sympathy groups respectively. Stiller and Dunbar [43] found that individual differences  in social cognition explained the size of an individual's support clique in a better way. On the other hand, the size of sympathy group is well explained by  individual's
performance on memory  tasks. Moreover, Dunbar [44] showed that six degrees  of separation did not necessarily mean that information would flow at uniform rate in one's social network. From [33] and [44], we can have the following question: do the trust levels remain same as of the
previous in case of one's stable relationships with 150
individuals in his networks if network's size starts to grow? This question is never asked in any literature of social networks with trust, network's size and Dunbar's number. In this paper, we answer to this question. For recent review on Dunbar's number, one may refer to [45].

\section{ Uniform distribution of trust}
In this section, we try to find nature of one's trust levels on acquaintances  of his social network when the network size starts to increase. \\

 Let us consider a social network with a population of $\textbf{N} ( > 1)$ individuals and
each of them has initial trust value $x\in  [0,1]$. We consider a constant $T_{c} \in [0,1]$ as the critical value of trust for an individual, above which he is susceptible to receive information as well as to transmit information to others. The individuals below the critical value of trust $T_{c}$ are considered to be non-reliable by their neighbors and thus they don't participate in
information transmission. Based on individuals'  roles in
information transmission, we divide the population into three categories. They are  given below. 
\begin{itemize} \item Susceptible (\textbf{S}) \item
Ignorant (\textbf{I}) \item Transmitters (\textbf{R})
\end{itemize}

The susceptible ($\textbf{S}$) people are those individuals who
can receive information and become a transmitter ($\textbf{R}$).
The ignorant ($\textbf{I}$) category consists of individuals with
trust value below the critical value of trust $T_{c}$. We assume
$\beta$ as the probability with which a transmitter sends
information to the neighboring susceptible individuals. In this
paper, we are considering two types of distribution for trust on
individuals. One is uniform distribution and the other is
power-law distribution.\\

Now, we consider that all individuals are assigned trust values
which are selected from a uniform distribution in [0,1]. Let,
\textbf{f} be the fraction of people with trust value greater than
or equal to $T_{c}$. Hence, the effective number of susceptible
individuals is $S = N\textbf{f}$. In the case of uniform
distribution, the fraction of population who participate in
information diffusion is, $ \textbf{f} = 1-T_{c} $ and the
fraction of non-reliable or ignorant population is, $ 1-\textbf{f}
= T_{c}$. Thus, dynamic equations regarding the information
diffusion in our case are given below.

\begin{equation}
    \frac{dI}{dt}= 0\\
\end{equation}
\begin{equation}
    \frac{dS}{dt}=\frac{-\beta SR}{N} \\
\end{equation}
\begin{equation}
    \frac{dR}{dt}=\frac{\beta SR}{N} \\
\end{equation}

To normalize the dynamics equations, we
consider $ \frac{S}{N} = \textbf{s} $. Hence, we obtain $
\textbf{s} = \textbf{f}$. Now, we consider $ \frac{R}{N} =
\textbf{r}$. So, at any instant t, we obtain $ \frac{I}{N} =
1-\textbf{f} -\textbf{r} $. If we denote $\frac{I}{N} =
\textbf{i}$, then $ \textbf{i} + \textbf{r}
+\textbf{f} = 1 $.\\

Thus, dynamic equations (1) to (3) are reduced to the following equations (4) to (6).
\begin{equation}
    \frac{d\textbf{i}}{dt}=0
\end{equation}
\begin{equation}
    \frac{d\textbf{f}}{dt}=-\beta \ \textbf{f}\textbf{r} \\
\end{equation}
\begin{equation}
    \frac{d\textbf{r}}{dt}=\beta \ \textbf{f}\textbf{r} \\\
\end{equation}

Since $\textbf{f} = 1- \textbf{i} - \textbf{r} $, thus, equation
(6) becomes
\begin{equation}
    \frac{d\textbf{r}}{\textbf{r}(1-\textbf{i}-\textbf{r})} = \beta 
dt
\end{equation}
We consider $\textbf{r}=r_{0}$ at $t=0$. As $\textbf{r}$ is time
dependent, hence the solution of (7) becomes
\begin{equation}
    \textbf{r}(t) = \left ( \frac{e^{(1-\textbf{i})\beta
t}(1- \textbf{i}) \ r_{0}}{1-\textbf{i} -r_{0} +
r_{0}e^{(1-\textbf{i})\beta t}} \right)
\end{equation}

It is a logistic solution of equation (7) with initial fraction of
transmitters as $r_{0}$. If $t \rightarrow\infty$, then $
\textbf{r}(t) \rightarrow 1 - \textbf{i}$. Thus, every susceptible
gets the information and becomes a transmitter.\\

When $\textbf{i} = 0$ at $t=0$, then the solution (8) reduces to
 \begin{equation}
    \textbf{r}(t)= \frac{r_{0}e^{\beta t}}{1-r_{0}+r_{0}e^{\beta t}} .
 \end{equation}
Thus, we obtain the  fraction of transmitters at any instance $t$
with initial fraction of transmitters $r_0$ in our social network
with $N(>1)$ individuals.
\section{Power-law distribution of trust}
According to Barab\'{a}si and Albert \cite{50}, large networks viz.  WWW, citation patterns in science, the collaboration graph of movie actors, etc. are scale-free networks. In a scale-free network, the probability $P(k)$ that a vertex in the network interacts with k other
vertices decays as a power-law i.e. $P(k)\propto k^{-\alpha}$. Prior to \cite{50}, Redner \cite{69} found the existence of power-law distribution in citations distribution of published papers having power-law exponent $\alpha \approx 3$. In order to generate a power-law distribution of trust with exponent $\alpha$ \ from a uniform distribution $P_{y}$ in [$x_{0}$,$x_{1}$], the random variable is given by, 
\begin{equation}
    X = \left[\left(x_{1}^{\alpha+1}-x_{0}^{\alpha+1} \right )y +
x_{0}^{\alpha+1}\right]^{\frac{1}{\alpha+1}}.
\end{equation}
It is the probability distribution function P(x) for generating
power-law with $x_{0}$ as the minimum value and $x_{1}$ as the
maximum value of x.  The minimum value of  $x$ cannot be zero for
a power-law distribution. So, we are fixing minimum value of $x$
as $0.1$ and maximum value of $x$ as $1$. Now, the minimum value
of  $x$ = $0.1$ means the minimum  value of trust in the
distribution cannot be lower than $0.1$. For the critical value of
trust $T_{c}$, in order to find  \textbf{f} in case of power-law
distribution, the fraction of individuals with trust values less
than $T_{c}$ is obtained and is subtracted from $1$.  The
probability distribution curve is obtained for $10^{6}$ values of
$y\in$ $[ 0.1,1 ]$ to get a smooth curve for power-law
distribution with power-law exponent $\alpha$. It is known that
for power-law distribution,  $2< \alpha< 3$. \ Using the data from
probability function of power-law and by fixing $T_{c}$, we can
obtain the initial fraction $\textbf{f}$ of population who are
susceptible with trust values  greater or equal to $T_{c}$. The
fraction \textbf{f} can be used for finding the number of
susceptible people in the information transmission by $
S=N\textbf{f}$.  Thus, fraction of people in the ignorant category
is $ \textbf{i} = 1 - \textbf{s} $. Remaining dynamics and equations are same as  of uniform trust
distribution.\\
\section{ Results and discussions}

This section consists of discussions regarding figure (1) to
figure (33). These simulations answer to our question which was
asked at the end of the introduction section.\\

From figure (1) to figure (3), we are considering a network of 150
(=N) individuals with uniform trust distribution. Information is
transmitted with transmission probabilities $\beta$=0.25 and
$\beta$=0.5 in figures (1) and (2) respectively with $R=1$ at
$t=0$, i.e. $R_{t=0} =1$. The total informed people level reaches
to Dunbar's hierarchy levels of 50, 15 and 5  at trust values
0.66, 0.90 and 0.96 respectively. In figure (2), the diffusion
occurs quicker with $\beta$=0.5 than that of $\beta$=0.25 in
figure (1). Figure(3) represents the total informed people for all
the cutoff values between 0 and 1 with an increment of 0.01 so as
to easily visualize Dunbar's hierarchy levels of stability and
required trust values. In case of figure (4) to figure(6), we are
considering a network of 500 (=N) individuals with uniform trust
distribution. Information transmission probabilities are same as
of figures (1) to (3) in figures (4) and (5). Now, we can see from
figure [6] that at trust values 0.7 and 0.9, the total informed
people level reach Dunbar's hierarchy levels of 150 and 50
respectively. Moreover, like network with $N=150$, the diffusion
of information occurs quicker at $\beta$=0.5 than that of
$\beta$=0.25 for networks with $N=500$.\\

Now, we consider a network with 1500 (=N) individuals  with
uniform trust distribution for figure (7) to figure(9). We can see
that at trust value 0.9, the total informed people level reaches
Dunbar's hierarchy level of 150. Like previous two networks,  it
can be concluded from  figures (7) and (8) that information
diffusion occurs quicker with $\beta$=0.5 than that of
$\beta$=0.25. As we extend our network size from 1500 to 5000 in
figure (10) to (12), we find from figure(12), trust level
increases to 0.97 to reach the stable connections of the network
i.e. Dunbar's number 150. Moreover, information diffusion property
remain same for $\beta=0.5$ as of our previous networks. From
above four network sizes, it is important to note that trust
levels remain same irrespective of information diffusion rate in a
network. It means that  trust level never helps to diffuse
information quickly or vice-versa to reach Dunbar's number 150
along with hierarchy layers of 5, 15 and 50 individuals in
networks of different sizes.\\

From figure (13) to figure (17),  we show that requirements of
trust levels increase for  the same individuals of one's  network
once the network size increases. From figure (13), to maintain the
trust on same 5 individuals, trust levels require 0.97 and 0.99
once the network size increases to 150 and 500 individuals
respectively. Similarly, to maintain the trust on same 15
individuals, trust levels require 0.90 and 0.97 once the network
size increases to 150 and 500 individuals respectively. Again from
figure (15), to maintain the trust on same 50 individuals, trust
levels require 0.66 and 0.90 once the network size increases to
150 and 500 individuals respectively. However, the trust levels
reach asymptotically to 1 if network size increases to  $\sim$5000
individuals. In case to maintain trust on same 150 individuals,
trust level increases from 0.70 for network of 500 individuals to
0.90 for network of 1500 individuals. Although one can not
maintain stable relationships  with more than 150 individuals, yet
increase of trust  level is observed  in figure (17). Thus, figure
(13) to figure (17)  conclude that requirement of trust level on
the same individuals of one's  social network increases if  the
size of the increases.  We are not sure about  conspiracies in
one's social network, but we hope that these increase of trust
levels as of increase in network sizes may be  a crucial reason to
study various factors  of several types of conspiracies or
deceptions against one by individuals of his social network.\\

Now, we turn our attention to power-law distribution of trust. In
case of power-law distribution of trust from figures (18) and
(19), if $\alpha=2.1$ and $R_{t=0}=1$ for network of size $N=150$,
then trust cutoff is $\sim$ 0.235 for 50 individuals for both
diffusion rates $\beta=0.25$ and $\beta=0.5$ to reach 50
individuals. Similarly from figures (21) and (22), if $N= 500$,
then trust cutoff is $\sim$ 0.26 to reach 150 individuals for
$\beta=0.25$ and $\beta=0.5$. For figures (24) and (25), if $N=
1500$, then trust cutoff is $\sim$ 0.50 to reach 150 individuals
for $\beta=0.25$ and $\beta=0.5$. Again for figures (27) and (28),
if $N= 5000$, then trust cutoff is $\sim$ 0.79 to reach 150
individuals for $\beta=0.25$ and $\beta=0.5$.  Thus, trust
levels for same 150 individuals increase with respect to increase
in the  size of the network independent to diffusion rate.
Therefore, one cannot expect that an information will be shared by
same individuals if his network size increases.\\

In case of variations of total informed people in one's network
with size $N=150$, the trust cutoff increases to reach 50
individuals for figure (20) with decrease in values of power-law
exponent $\alpha$, where $2< \alpha <3$. For figure (20), trust
cutoffs are 0.175 and 0.240 for  $\alpha = 2.9$ and $\alpha = 2.1$
respectively. Similar result can be observed to reach 150
individuals for same values of power-law exponents in figures
(23), (26) and (29) but with different sizes of networks. 
    As observed from the simulation, we can infer that the trust cutoffs increase with decrease in the power-law exponents $\alpha$ i.e. $\alpha$ $\propto$ ${1}/$({{Trust} {cut off}}). Thus, we have the following conjecture from  the above discussions.\\

{\bf Conjuncture:} Conspiracies or deceptions are caused in one's social network if he doesn't increase his levels of trust on individuals of his social networks  once size of his social network increases.\\

\section{Comparison with  online scenarios in Facebook}
This section compares some of our above findings to recent experimental results related to online social groups. In 2014, Kraut and Foire \cite{46} surveyed  472,231 online groups of Facebook to find online groups' success and survival. They found that 57$\%$ of these groups  ceased to create new content by the end of three months. Groups were found  less likely to be survived when the founders had more ties to group members, when the founders invited more of the members and when the founders served as exclusive administrators of the group. Moreover, larger groups  which engaged founders vigorously in first week of the formation of those online groups were found to be more successful than smaller and less engaging groups. Similar to above, Ma et al. \cite{47} surveyed 6,383 Facebook groups users about their trust attitudes. Following aggregated  behavioral and demographic data for group members, they concluded that (1) an individual's propensity to trust is associated with how they trust their groups, (2) groups which are smaller, closed, older, more exclusive, or more homogeneous groups are trusted more, and (3) trust can be predicted from a group's overall friendship-network structure and an individual's position within the structure. In \cite{47}, Ma et al. also found that trust decreases with the increase of group size. But interestingly, they confirmed the finding of Kraut and Foire \cite{46} that trust grows with the number of group administrators.  They also suggested that less popular groups must be recommended to individuals because of diversity of group recommendations, trust and user satisfaction.  Ma et al. \cite{47} did not find any significant difference between  closed and secret groups once the groups' sizes  crossed Dunbar's number 150. Moreover, Cohen and Havlin \cite{48} and  Guardiola et al. \cite{49} found that trust cooperation networks posses scale-free degree distributions. Since, power-law exponents $\alpha$ have major roles in  scale-free degree distributions, thus  it can be understood from \cite{46}, \cite{47}, and \cite{48}  that our result  $\alpha$ $\propto$ ${1}/$({{Trust} {cut off}}) may have significant impact in online and offline social networks. Again, preferential attachments are present in scale-free networks \cite{50, 51}, thus it can be found that number of individuals with less trust cutoffs in an individual's social network increases if the size of his social network grows. So, trust cutoff is not absolute in social network. Hence, one must have to increase  trust cutoffs on individuals of his social network if he doesn't want to be deceived by members of his social network due to increase of the size of his social network. Thus, we have found that one of our predictions matched with one of the data based observations  related to trust and size of social networks of  Ma et al. \cite{47}. Thus, we hope to have more rigorous data based results to validate our claims.\\

\section{Conclusion}
In this work we tried to find out how various types of trust distributions like uniform and power-law in a network results in the information diffusion among them as well as how the change in population make the difference in the trust level of individuals to keep the same number of friends in each hierarchy level of their friendship. We found that trust levels remain same irrespective of information diffusion rate in a network. It means that  trust level never helps to diffuse information quickly or vice-versa to reach Dunbar's number 150 along with hierarchy layers of 5, 15 and 50 individuals in networks of different sizes. Moreover, requirement of trust level on the same individuals of one's  social network increases if  the size of his social network increases. Also, trust levels for same 150 individuals increase with respect to increase in the  size of the network independent to diffusion rate. Therefore, one cannot expect that an information will be shared by same individuals if his network size increases.  We can infer that the trust cutoffs increase with decrease in the power-law exponents $\alpha$ i.e. $\alpha$ $\propto$ ${1}/$({{Trust} {cut off}}). However we are not sure whether conspiracies or deceptions are caused in one's social network if he doesn't increase his levels of trust on individuals of his social networks  once size of his social network increases. Thus, we make it as a conjecture with a hope that we will have the validity of the conjecture in the future.\\ 

\section{Open questions}
In this section, we propose following questions which are open till now in research related to social networks and trust.\\

{\bf Q.1.} What are the topological and graph theoretical features of one's social network in terms of centrality, connectivity, etc. if trust levels are increased on  members of the network. \\

{\bf Q.2.} Are there  hidden mathematical and social networks patterns available in wars  which are caused due to deception?\\

{\bf Conflict of  interest:} The authors declare that there is  no conflict of interest.\\

{\bf Acknowledgement:} This research was supported in part by the International Centre for Theoretical Sciences (ICTS) during a visit of the first and second authors for participating in the program-Summer research program on Dynamics of Complex Systems (Code: ICTS/Prog-DCS2019/07). The authors are thankful to Prof. Robin I. M. Dunbar of University of Oxford for his constructive discussions during the preparation of this paper.

\newpage
\begin{figure}[h!]
    \centering
    \includegraphics[width=12cm]{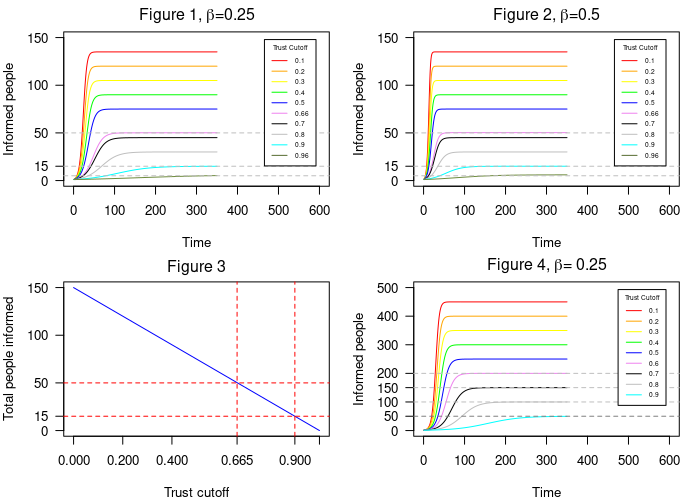}
    \label{fig:my_label}
\end{figure}
\begin{figure}[h!]
    \centering
    \includegraphics[width=12cm]{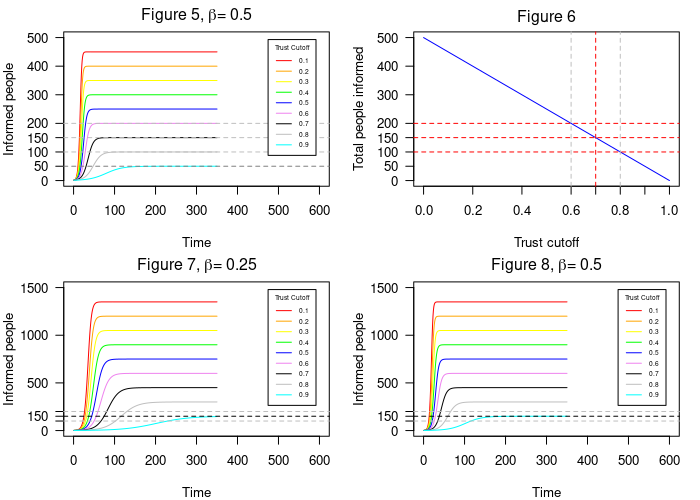}
    \label{fig:my_label}
\end{figure}
\begin{figure}[h]
    \centering
    \includegraphics[width=12cm]{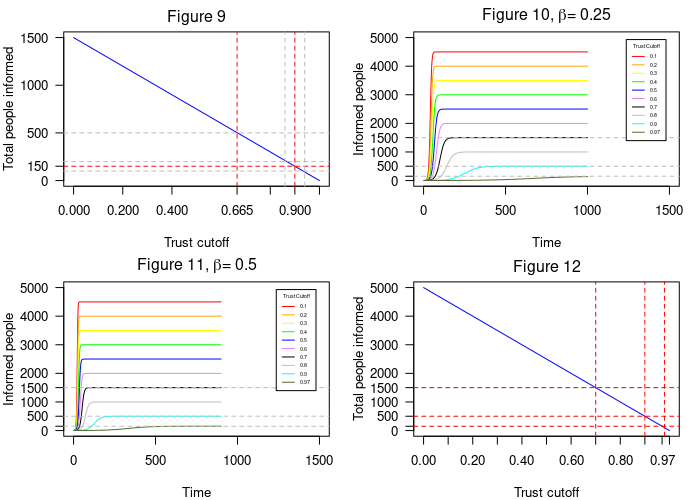}
    \label{fig:my_label}
\end{figure}

\begin{figure}[h]
    \centering
    \includegraphics[width=12cm]{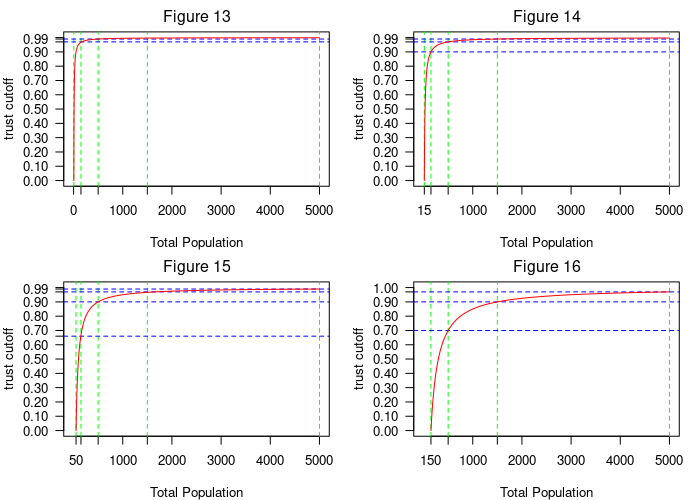}
    \label{fig:my_label}
\end{figure}
\begin{figure}[h]
    \centering
    \includegraphics[width=12cm]{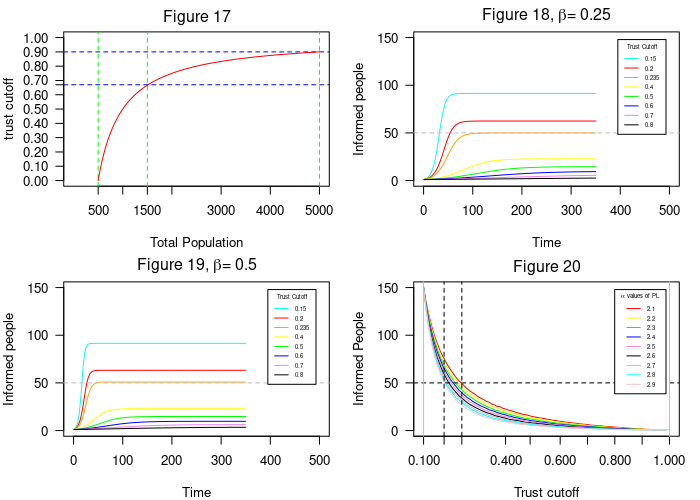}
    \label{fig:my_label}
\end{figure}
\begin{figure}[h]
    \centering
    \includegraphics[width=12cm]{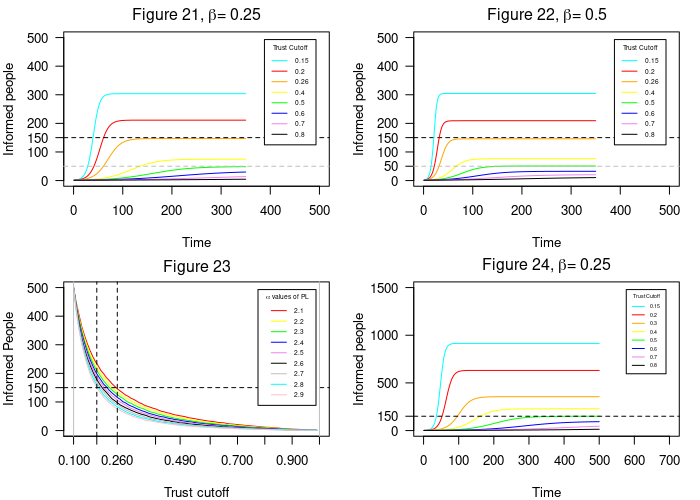}
    \label{fig:my_label}
\end{figure}
\begin{figure}[h]
    \centering
    \includegraphics[width=12cm]{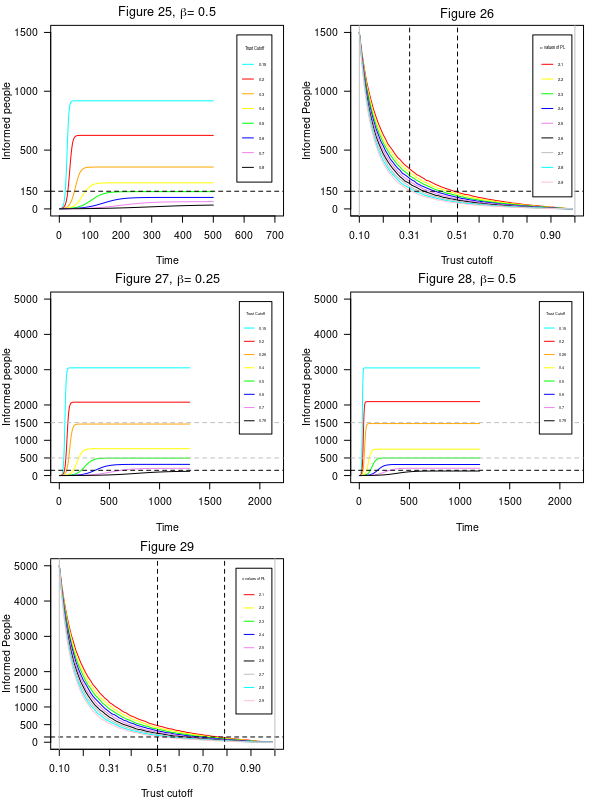}
    \label{fig:my_label}
\end{figure}

\end{document}